\documentclass[twocolumn]{el-author}

\usepackage{algorithmic}
\usepackage{algorithm}
\usepackage{graphicx}

\begin{document}

\title{An Efficient Local Strategy to Control Information Spreading in Network}

\author{Anoop Mehta and Ruchir Gupta}

\abstract{In social networks, control of rumor spread is an active area of research. SIR model is generally used to study the rumor dynamics in network while considering the rumor as an epidemic. In disease spreading model, epidemic is controlled by removing central nodes in the network. Full network information is needed for such removal. To have the information of complete network is difficult proposition. As a consequence, the search of an algorithm that may control epidemic without needing global information is a matter of great interest. In this paper, an immunization strategy is proposed that uses only local information available at a node, viz. degree of the node and average degree of its neighbour nodes. Proposed algorithm has been evaluated for scale-free network using SIR model. Numerical results show that proposed method has less complexity and gives significantly better results in comparison with other strategies while using only local information.}

\maketitle

\section{Introduction}
Spread of information in complex network and its control has been extensively studied in recent past. A few researchers have tried to expedite the dispersion by gossip \cite{1}, spread maximization \cite{2} etc, while others have tried to stop it \cite{4}, \cite{5}. Generally, a viral information or rumor is modeled as an epidemic and its dynamics is studied using the famous compartmental model in epidemiology, SIR (susceptible-infected-resistant) model \cite{3}. In disease dynamics, if a person comes into contact with an infected person then the person becomes infected with some probability. To control the infection spreading, immunization is required. To immunize every node is a difficult and expensive process. Therefore, it is imperative to find methods that may stop the spread by immunizing minimum number of nodes while allowing disease to spread to as few as possible. Substantial work that has been done on this regard can be found in \cite{4}, \cite{5}, \cite{6}.

Central node (the node that affects more nodes) identification is the key to stop the spread this is because the central node can pass infection to more nodes. Hence, if we can find out most central nodes in the network and immunize them, spread of disease can be controlled. Immunization of central nodes may keep more number of nodes away from infection. To identify the most central nodes, there are different kind of centrality measures defined in the literature viz. degree centrality, betweenness centrality, structural centrality, community centrality etc.. Computation of degree centrality, betweenness centrality, structural centrality needs global information of network, whereas community centrality needs the local information available in a community. Generally it is difficult to have the global information therefore the centrality method that needs only local information is preferred.

Degree centrality is defined as the number of connections a node has. Higher the number of connections, more central the node is considered \cite{7}. In degree centrality based immunization, the nodes with highest degree will be immunized. Although the method is simple, it requires the information of whole network, and we will later show that it is not always optimal to remove the highest degree node. Betweenness centrality defined as the ratio of number of shortest path passing through particular node versus the total shortest path between all pairs of nodes, is also based on global information \cite{8}. The main drawback of the strategy is its higher complexity. Structural centrality measures which node is close to center according to the structure of the network \cite{9}. This strategy is not reliable for large network due to high complexity and it also requires global information.

In community centrality, which is based on local information, identifies communities in the network and immunization is done to the nodes that are more central in a community \cite{5}. Results of community based centrality are comparable with degree centrality. Our proposed method is also based on local strategy. It runs on every node of the network simultaneously. For running the proposed algorithm a node only requires its own degree and average degree of neighbouring nodes. In the experiment each infected node can infect neighbouring susceptible nodes up to 5 rounds after that it cannot spread infection even if it is infected node. Proposed method only needs a very small number of nodes (4{$\%$} of the nodes) to be immunized for stopping the spread. The complexity of the proposed method is also less.

\section{Epidemic Model}
We have used the SIR model for scale-free networks to observe the performance of proposed and existing methods. SIR is taken to model epidemic dynamics of the network where each node is in one of the three states: node that have not received infection so far i.e. susceptible (S), node that have received infection and still in infected state i.e. infected (I) and node that have received infection and now recovered from infected state, i.e. resistant (R) and it cannot be infected again.

In this model, the infection starts with few infected and remaining susceptible nodes, infected nodes spreads the infection to neighbouring susceptible nodes with probability {$\lambda$}, we call it spreading rate and infected nodes become resistant with probability {$\sigma$}, we call it recovery rate. In the experiment we are taking {$\lambda$} = 0.1 and {$\sigma$} = 0.1. SIR model is defined as:
\newline

\begin{center}
	{$I + S \xrightarrow{\lambda} I + I$}
	
	{$I \xrightarrow{\sigma} R$}
\end{center}

\section{Network Model}
The commonly used LFR benchmark algorithm  \cite{10} is used to generate network. It generates a scale free network based on configuration model with power law degree distribution (more number of nodes with high degree and less number of nodes with low degree). It requires parameter like number of nodes(n), average degree{($k_{avg}$)}, maximum degree{($k_{max}$)} and exponent {$\gamma$} {$(1 < \gamma \leq 3)$}.

\section{Proposed Strategy}
Proposed algorithm may be divided in two steps. In first step node simultaneously computes the centrality of its neighbouring nodes and then in second step the node with highest centrality among the neighbours is immunized if found infected. If a node is immunized, the immunized nodes and its neighbours will stop the process. Proposed centrality is defined as:

{$$NC(i) = \frac{d_i}{(\sum_{j \in N_i}{d_j})/{d_i}}$$}

Here {$NC(i)$} is the centrality of node {$i$}, {$d_i$} is the degree of node {$i$} and {$N_i$} is the set of neighbours of node {$i$}.

Proposed centrality is the ratio of degree of a node and average degree of its neighbouring nodes. A node is considered more central if the ratio is high. Algorithm 1, describes how proposed strategy work in the network.

\begin{algorithm}[H]
 	\caption{$Proposed Immunization Strategy$}
 	\label{alg1}
 	\begin{algorithmic}[1]
 		\renewcommand{\algorithmicrequire}{\textbf{Input:}}
 		\REQUIRE {$N_i$}: Set of neighbours of node {$i$}
 		
 		\FOR{Node {$i$}}
 		\FOR{$j \in N_i$}
 		\STATE $d_{sum}(j)$ = 0;
 		\FOR{$k \in N_j$}
 		\STATE $d_{sum}(j)$ = $d_{sum}(j)$ + $|N_k|$
 		\COMMENT {Total degree of neighbouring nodes}
 		\ENDFOR
 		\STATE $d_{avg}(j)$ = $d_{sum}(j)$/$|N_j|$
 		\COMMENT {Average degree of neighbouring nodes}
 		\STATE $NC(j)$ = $|N_j|$/$d_{avg}(j)$
 		\COMMENT {{$NC(j)$}: Centrality of {$j^{th}$} node}
 		\ENDFOR
 		\IF{(V[i] != 0)}
 		\STATE t = max [$NC(1)$,...,$NC(|N_i|)$]
 		
 		\COMMENT {Highest centrality node among neighbouring nodes of {$i^{th}$} node}
 		\IF{t is infected node}
 		\STATE Immunize node t
 		\STATE V[t] = 0
 		\STATE V[k] = 0, where k is neighbours of t
 		
 		\COMMENT {Now onwards t and its neighbours will stop the immunization process}
 		\ENDIF
 		\ENDIF
 		\ENDFOR
 		
 	\end{algorithmic}
\end{algorithm}
 
Proposed method only requires the information that is available with its neighbouring nodes and node itself. No global information is required. Proposed centrality identifies the degree of influence that a node has in its neighbourhood and therefore with the help of this centrality, most influence node in a particular region can be identified contrary to degree centrality that does not take the vicinity into account. To understand the effectiveness of the method, let us consider an extreme case where a graph has two components, a complete subgraph with {$n$} nodes and a star subgraph with the center node of degree {$n-1$}. For epidemic control, degree centrality  would immunize any node from the complete subgraph or the center node of the star subgraph. But it is more effective to remove the infected center node of the star than to immunize an infected node from the complete graph, since in the former case the nodes in the star graph would be isolated from the infection, while in the later removing a node from a complete graph still keeps the nodes connected for infection spreading.
 
Removal of high ratio infected nodes creates isolated components that are disconnected for infection spreading. These isolated components eventually covers the complete work. So as we keep immunizing centrally infected nodes, the network will be decomposed into disease free components and infection spreading would stop.

\section{Experimental Results}
For performance measurement and comparison with existing method, networks are generated using LFR algorithm. We have used parameters like number of nodes(n) = 5000, maximum degree({$k_{max}$}) = 120, average degree({$k_{avg}$}) = 8 and exponent {$\gamma$} = 3 to generate networks.

To study epidemic spreading model using SIR model, initially we have taken 1{$\%$} of nodes as infected and remaining nodes are susceptible nodes. Each infected node can infect neighbouring susceptible nodes up to 5 rounds after that it cannot spread infection even if it is infected node.

The experiment is performed in three steps for each round, in first step infection spreads, in second step infected nodes recover with recovery rate and in third step fraction of nodes are immunized. In existing methods, third step happens only once and remove a fraction of nodes in each experiment, but in proposed local strategy some fraction of nodes are removed in every round. The experiment is assumed to be completed when there is no infected node left in the network. Finally the total number of infected nodes in whole experiment is observed. We have observed the results with the removal of different fraction of nodes in existing methods. Each experiment is conducted on 10 different networks generated using the same parameter. Mean all 10 results has been shown in figures.

\begin{figure}[!htb]
	\minipage{0.24\textwidth}
	\includegraphics[width=\linewidth]{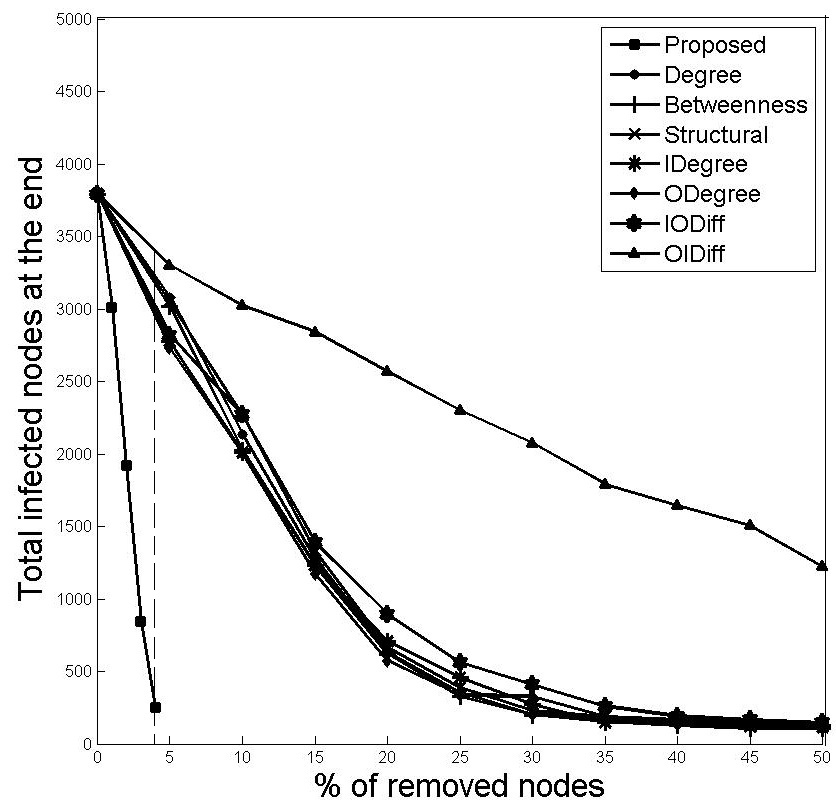}
	\centering
	(a)
	\label{fig.1}
	\endminipage\hfill
	\minipage{0.24\textwidth}
	\includegraphics[width=\linewidth]{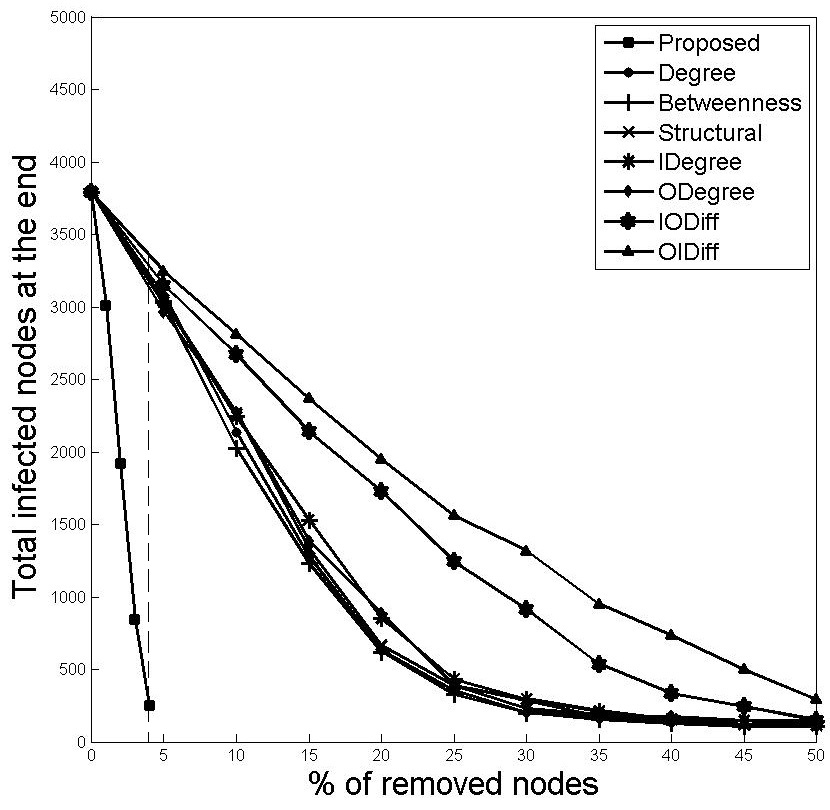}
	\centering
	(b)
	\label{fig.2}
	\endminipage
	\newline
	%
	\minipage{0.24\textwidth}
	\includegraphics[width=\linewidth]{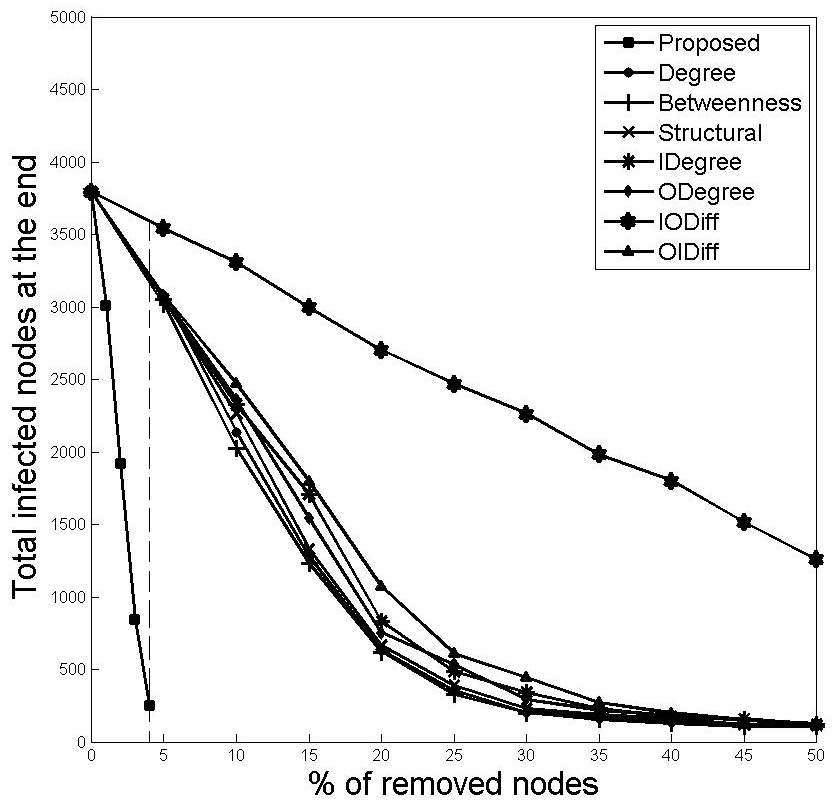}
	\centering
	(c)
	\label{fig.3}
	\endminipage\hfill
	\minipage{0.24\textwidth}
	\includegraphics[width=\linewidth, height=41mm]{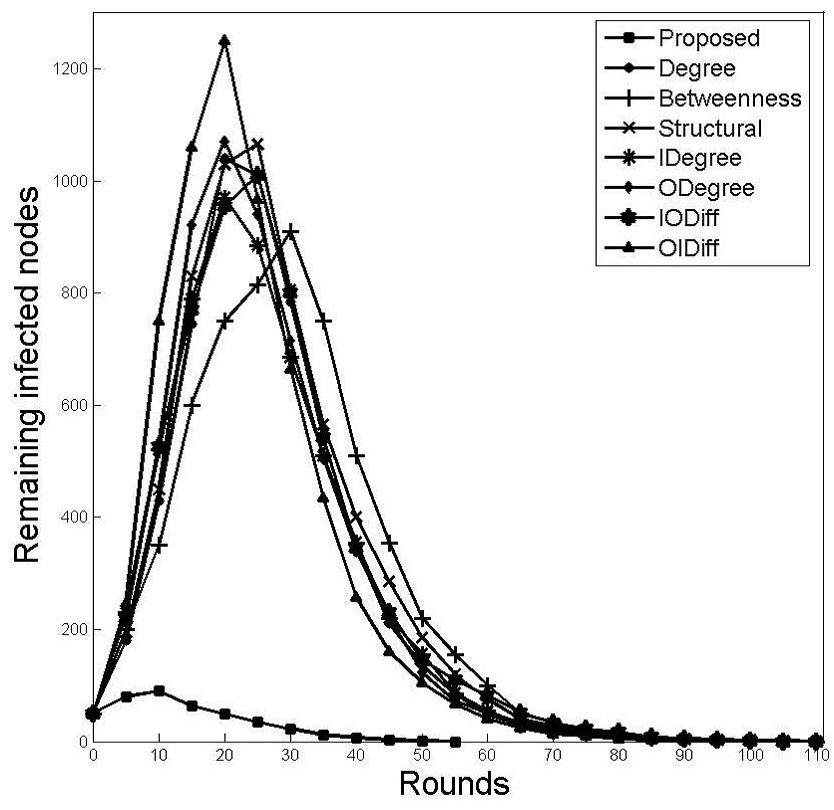}
	\centering
	(d)
	\label{fig.4}
	\endminipage
	\caption{Effect of various immunization strategies with {$\lambda$} = 0.1, {$\sigma$} = 0.1 and (a) {$\mu$} = 0.3 (b) {$\mu$} = 0.5 (c) {$\mu$} = 0.7 (d) Time evolution of infection for 4{$\%$} immunization with {$\mu$} = 0.3}
\end{figure}

We have plotted the total number of infected at the end versus the percentage of nodes removed for different values of {$\mu$} in fig. 1(a), 1(b) and 1(c) for 0.3, 0.5 and 0.7 respectively. The value of {$\lambda$} and {$\sigma$} is taken as 0.1 and 0.1 respectively for all the three figures. Here it is important to note that inter community links increases with the increase in value of {$\mu$} in the network.

Experimental results show that, proposed algorithm needs to immunize only 4{$\%$} of the nodes to stop infection spreading, whereas other strategies requires to immunize around 30{$\%$} or more nodes to stop the spread. In fig. 1(a), 1(b) and 1(c) dotted lines shows total infected nodes when 4{$\%$} of nodes are immunized, because in proposed method we require only 4{$\%$} immunization to stop spread. So in fig. 1(d) we have shown the effect at 4{$\%$} immunization by the number of infected nodes remaining in each round for different strategies. It is evident from the figure that initially infection spreads but after some rounds it starts to decrease and stop spread. Proposed centrality stops infection spreading after 30 rounds whereas existing centralities requires around 70 rounds for stopping spread.

Table 1 shows the time and space complexity of different algorithms. Here {$n$} is the total number of nodes in the network and {$c$} is the highest nodal degree in the network. It can be seen from the table that the space complexity of the proposed algorithm is either less or comparable with existing algorithms whereas time complexity is higher than degree centrality and less than all other algorithms.

\begin{table}[!t]
	\begin{center}
		\begin{tabular}{ |c|c|c| } 
			\hline
			 & Time Complexity & Space Complexity \\ 
			\hline
			Degree Centrality & {$O(n)$} & {$O(n)$}\\ 
			Betweenness Centrality & {$O(n^3)$} & {$O(n^2)$}\\
			Structural Centrality & {$O(n^2)$} & {$O(n^2)$} \\ 
			Community & {$O(kn)$} & {$O(n)$} \\ 
			Proposed Centrality & {$O(c^2)$} & {$O(n)$} \\ 
			\hline
		\end{tabular}
	\end{center}
	\caption{Time and space complexity comparison of centrality}
	\label{tabl1}
\end{table}

\section{Conclusion}
For large network global strategy is not optimal because it requires global information of network to control infection spreading. The proposed local strategy control infection spreading using local information at neighbouring nodes. Experimental result show that it immunize less number of nodes compare to other strategies and complexity of the proposed neighbour centrality is also less.

\end{document}